\documentclass[%
 reprint, superscriptaddress,
 amsmath,amssymb,
 aps,
]{revtex4-2}

\usepackage{graphicx}
\usepackage{dcolumn}
\usepackage{bm}
\usepackage{xcolor}

\begin{document}

\preprint{APS/123-QED}

\title{A Tactile Void}

\author{P. Tapie}
\address{Laboratoire Jean Perrin, UMR 8237 Sorbonne Université/CNRS, Institut de Biologie Paris Seine, 4 Place Jussieu, F-75005 Paris, France}
\author{D. Barreiros Scatamburlo}%
\address{Laboratoire Jean Perrin, UMR 8237 Sorbonne Université/CNRS, Institut de Biologie Paris Seine, 4 Place Jussieu, F-75005 Paris, France}
\author{A. Chateauminois}
\address{Laboratoire Sciences et Ingénierie de la Matière Molle, UMR 7615 CNRS, ESPCI Paris, Université PSL, Sorbonne Université, F-75005 Paris, France}
\author{E. Wandersman}
\email{elie.wandersman@sorbonne-universite.fr}
\address{Laboratoire Jean Perrin, UMR 8237 Sorbonne Université/CNRS, Institut de Biologie Paris Seine, 4 Place Jussieu, F-75005 Paris, France}

\date{\today}

\begin{abstract}
We mimic the mechanical response of touch mechanoreceptors by that of a gas cavity embedded in an elastic semi-cylinder, as a fingertip analogue. Using tribological experiments combined with optical imaging, we measure the dynamics and deformation of the cavity as the semi-cylinder is put in static contact or slid against model rough surfaces at constant normal force and velocity. We propose an elastic model to predict the cavity deformation under normal load showing that membrane mechanical stresses are anisotropic and we discuss its possible biological consequences. In friction experiments, we show that the cavity shape fluctuations allow for texture discriminations.

\end{abstract}

\maketitle
\section{Introduction}
Touch mechanoreceptors are neural cells embedded in the dermis of mammals~\cite{deflorio2022skin,handler2021mechanosensory,johansson1979tactile}, implied in the sense of touch. In humans the density of mechanoreceptors is particularly high around the fingertip regions and in the palm of the hand~\cite{johansson1983tactile,vallbo1984properties} and enables an exquisite tactile sensitivity as we  explore a solid surface with our fingers, to probe its shape or roughness. Upon application of contact stresses, the skin is deformed and mechanical stresses are conveyed from the surface of the skin to the embedded mechanoreceptors, which transform mechanical signals into neural signals propagating toward the central neural system. This mechanotransduction process is performed, at the microscopic scale, by mechanosensitive transmembrane proteins (such as the Piezo protein familly~\cite{ranade2014piezo2,woo2014piezo2,zhao2018structure,wang2019structure,delmas2022piezo}) inserted in the plasmic membrane of mechanoreceptors. The structure of these mechanosensitive proteins, and as a consequence its ion permeability depend on the membrane stresses. Under a mechanical stress, an ion flow can be triggered across the membrane, yielding to an electrical depolarization of the cell, this is the birth of an action potential.\\
\par The physiology of mechanoreceptors is well documented~\cite{handler2021mechanosensory} as well as their neural response, using micro-neurography experiments~\cite{knibestol1970single,johansson1976skin}. In particular, two different classes of mechanoreceptors have been identified~\cite{handler2021mechanosensory} : Slowly Adapting (SA) mechanoreceptors on the one hand, whose neural response is triggered during the whole time duration of a mechanical stimuli and Fast Adapting (FA) mechanoreceptors, with a neural response that depends on the time variations of the stimuli~\cite{bolanowski1988four}. Even if the physiology and protein sequence and structure of some mechanosensitive proteins have been recently identified, it is yet not clear to state how they precisely encode mechanical signals, and what sets microscopically peculiar dynamical responses of FA mechanoreceptors.  Even from a purely mechanical perspective, the way membrane stresses are distributed (angularly and temporarily) during a typical tactile exploration have not been measured nor modeled. \\
\par  Prior to any neural filtering, there is indeed a first purely mechanical filtering of the tactile information, performed by the geometrical and mechanical properties of the tactile organ. For instance, the presence of fingerprints (epidermal grooves with a typical wavelength $\lambda$) at the extremity of the fingers modulates the subcutaneous stresses~\cite{scheibert2009role, wandersman2011texture} at a temporal frequency $v/\lambda$, where $v$ is the exploration velocity. In addition, the resonance properties of the skin have been shown to participate into tactile perception, by propagating mechanical stresses on large distances and thus triggering more mechanoreceptors response, even outside the finger/surface contact zone~\cite{manfredi2012effect}. But beside these macroscopic mechanical filtering processes, one may wonder how the mechanoreceptor structure \emph{itself}, as a cellular inclusion in the extracellular matrix encode mechanical stresses, this is the purpose of this Letter.\\
 In this work, we use a biomimetic approach and model the mechanoreceptor by a gas cavity embedded in an elastomer semi-cylinder, as a fingertip analogue. We use optical imaging to measure the dynamics and deformation of the cavity as the finger is slid at constant velocity and normal force against a model rough surface. We discuss how the roughness of the surface is encoded in steady and fluctuating stresses at the cavity surface.\\
 \begin{figure*}
\includegraphics[width=\textwidth]{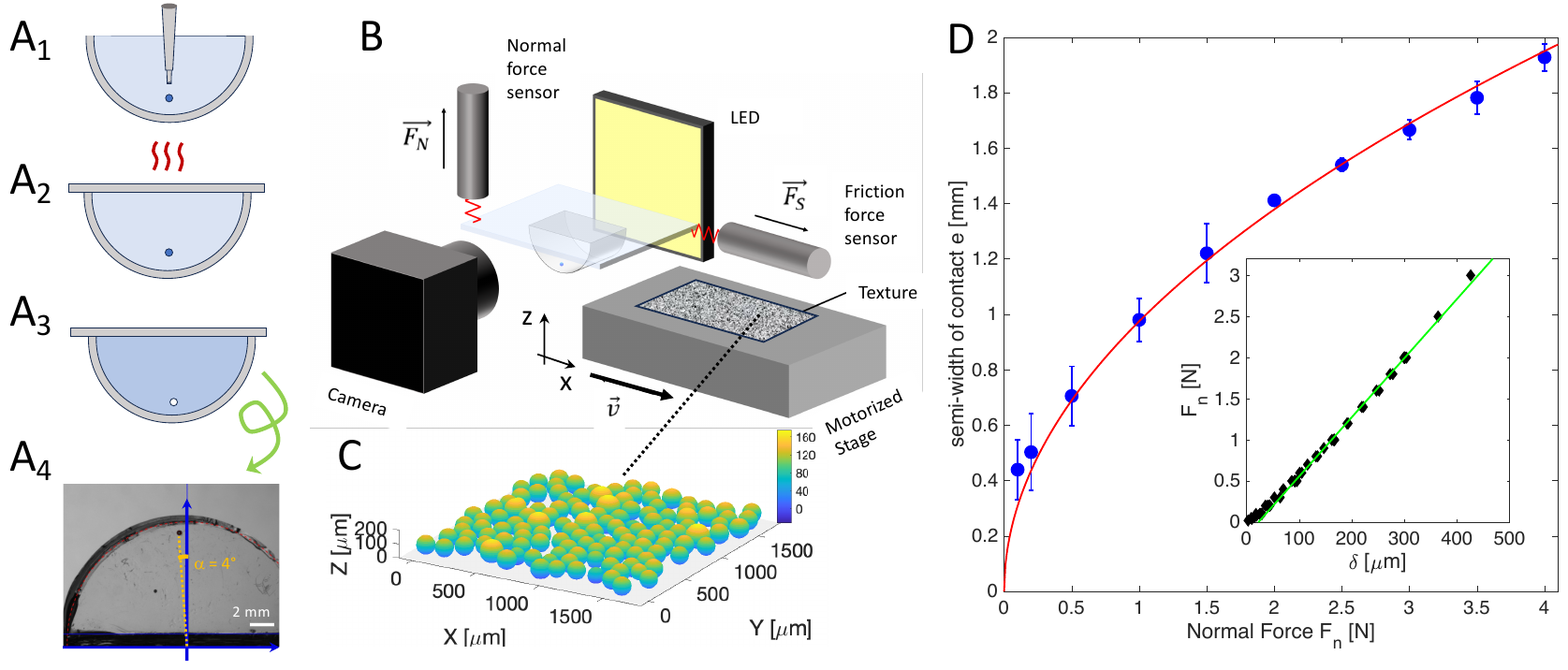}
\caption{\label{Fig1} A1-A3) Sketch of the artificial finger/mechanoreceptor fabrication. The unreacted PDMS mix is poured in a hemi-cylindrical Plexiglas mold. An aqueous droplet ($V\approx 0.5 \mu$L) is placed near the cylinder apex. Upon reticulation of the PDMS in an oven, the aqueous droplet evaporates and leaves a gas cavity in the elastic semi-cylinder.  A4) Macroscope side-view image of the PDMS semi-cylinder showing the location of the cavity close to the apex. B) Sketch of the experimental setup.  C) Z-profile of the rough- surface, from profilometry measurements (see Supplementary Materials). D) Semi-width of contact $e$ as a function of the normal force. The solid line is a fit by a Hertz cylinder/plane model (see text). Inset : Normal force as a function of the cylinder indentation $\delta$. The solid line is a linear fit (see text).}
\end{figure*}
 
 \section{Artificial finger and mechanoreceptor}
  The biomimetic finger is an elastic semi-cylinder of Poly(dimethylsiloxane) (PDMS) elastomer, using a 50/50 \%wt mix of Sylgard 184 and Sylgard 527 (Dow Corning inc.). The mix is vigorously stirred, degassed with first a centrifugation phase (3500 rpm, 10 min) and then let in a vacuum chamber for 30 min. This liquid PDMS mixture is placed in a Plexiglas hemi-cylindrical mold (\textit{see} Fig.~\ref{Fig1}A1-A3, inner radius $R=10$~mm, inner height $h$=~10 mm and length $L$=~10 mm). One droplet of deionized water (volume $\sim 0.5 \mu$L) is carefully added to the mix with a micropipette, and positioned at the proximity of the cylinder apex. A plexiglas coverslip is added at the flat top of the hemi-cylindrical mold and the system is placed in an oven at 65\textdegree C for 2 hours for PDMS cross-linking. During crosslinking, we observed that the water of the droplet evaporates through the permeable PDMS rubber, yielding after unmolding a gas spherical cavity of radius $a=224\pm1\mu$m embedded in the PDMS elastomer, at a position $z\approx$ 1 mm from the cylinder apex (\textit{see} Fig.~1~A4). The semi-cylinder rectangular base is glued to a plasma-cleaned glass plate using a few drops of PDMS which are crosslinked for a few minutes. The glass plate is then mounted on a tribological setup (\textit{see} Fig.~\ref{Fig1}B) very similar to the one used in \cite{wandersman2011texture}. Briefly, the glass plate holder is mounted on a two crossed dual cantilever beams (whose stiffnesses have been independently measured). The deflections of the two cantilevers are measured with two capacitive sensors, from which normal $F_n$ and tangential forces $F_s$ are deduced (range 0-2N), with measurement noises of about 50 mN and 10 mN, respectively. The semi-cylinder apex is positioned over a rough surface mounted on a manual $Z$ translation stage (to indent the finger) and a X/Y motorized stage (ICLS-200, Newport inc.), allowing to slide the surface at constant velocity  ($v$, from 0.05 to 0.2 mm/s). We used two different rough surfaces. The less rough one ($rough^-$) is made of poly(styrene) microspheres (diameter $d$=140~$\mu$m) spread and glued on a microscope glass slide the surface of which has been spin-coated with epoxy glue (Fig.~1C). The rougher one ($rough^+$)  is made using polydisperse glass microspheres ($d$= from 125 to 600 $\mu$m) glued the same way. The roughness of the rough surface is measured using a optical profilometer (Zegage Pro, Ametek inc, objective magnification x5) taking images over few tens of locations in each sample (see details in Supplementary Materials), from which we extract asperity height fields $h(x,y)$. We found respectively mean ($\pm$ rms) asperity heights of  141$\pm$16 $\mu$m for $rough^-$ and 325 $\pm$ 142 $\mu$m for $rough^+$ (See Supplementary Materials). Last, the cavity is imaged in transmission with a monocular zoom lens equipped with a PointGrey BlackFly S camera (1280x1024 pix$^2$) with a 5x magnification objective, yielding an image spatial resolution of  1.7 $\mu$m/pixel. The Young's modulus $E$ of the PDMS elastomer was measured using two methods, first by measuring the contact semi-width $e$ as a function of the normal force, as the finger is indented against a smooth glass surface and fitting the data with a Hertz contact model for an incompressible body~\cite{johnson1987contact} ($e=\sqrt{3F_nR/(\pi L E)} $, see Fig.1D), secondly by fitting linearly the normal force/indentation ($F_n/\delta$) relationship (inset of Fig.~1D) to an approximate solution $F\approx \pi L E \delta/3$ \cite{popov2010contact}. From an average of the two measurements, we obtain $E\approx 0.8\pm 0.2$ MPa, a value for PDMS mix in agreement with the results obtained in\cite{palchesko2012development}.\\

  \section{Results}
  \subsection{static contact}
\begin{figure}[t!]
\includegraphics[width=0.48\textwidth]{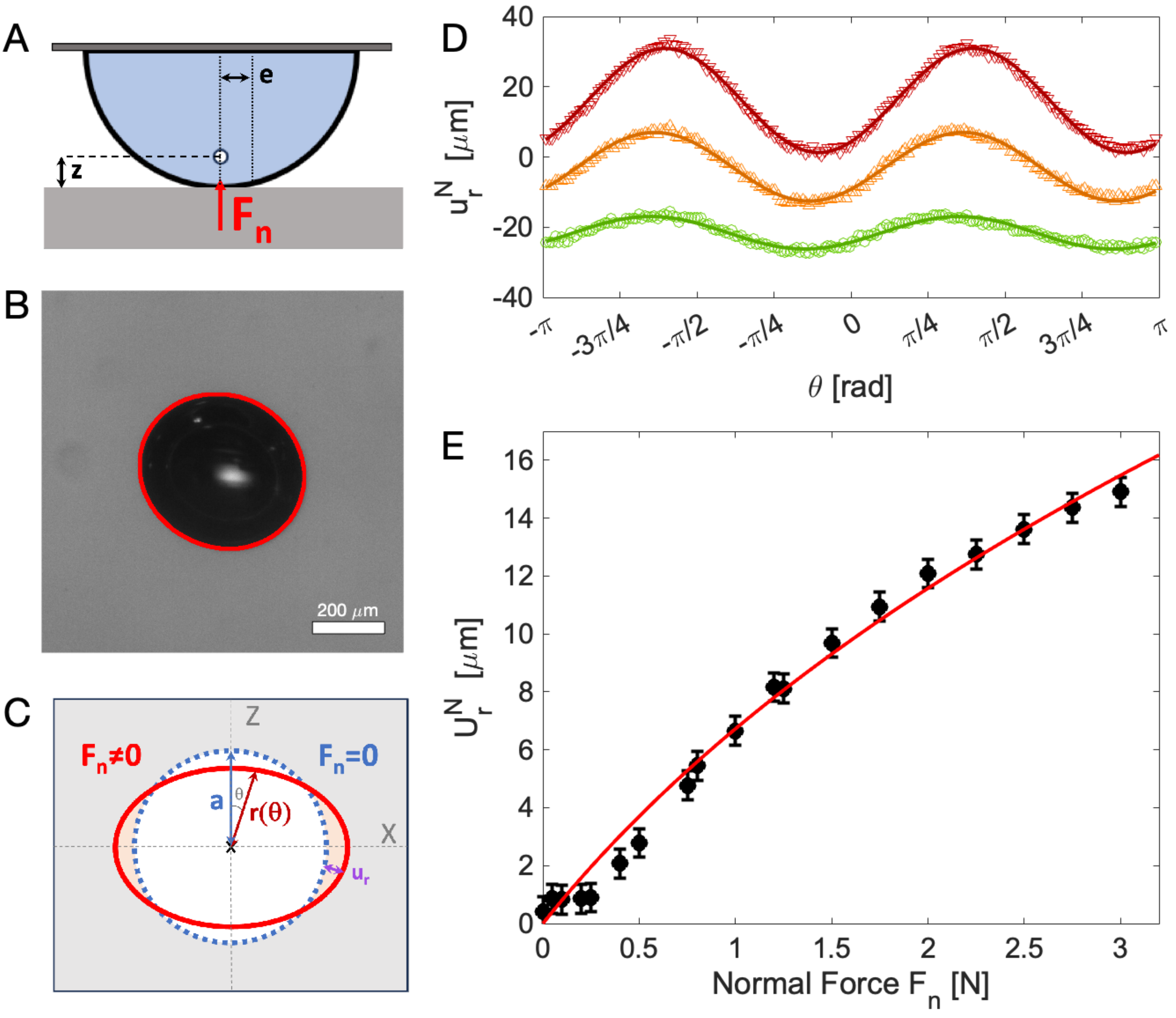}
\caption{\label{Fig2} A) Sketch of the semi-cylinder in static normal contact with a smooth plane. B) Image of the cavity under a $F=2.5$ N normal load. The detected contour has been superimposed. C) Sketch of the undeformed (dotted line) and deformed (dashed) cavity, from which we compute the radial displacement $u_r(\theta)$. D) Static radial displacement under normal load as a function of $\theta$, for normal force $F_n = 0.75$ N (lower curve), $F_n = 1.5$ N (middle curve), $F_n = 3$ N (upper curve). These curves have been shifted vertically arbitrarily for sake of clarity. The datas are adjusted by $u_r = A+ U_r^N \cos{(2\theta+\varphi)}$ (solid lines). E) Radial displacement amplitude $U_r^N$ as a function of the normal force. The measurement noise of 1$\mu$m is obtained from the standard deviation of $U_r^N$ of repeated out-of-contact experiments. The solid line is a fit with Eq.\ref{sigma_cylindre}}.  
\end{figure}

  We first performed \emph{static} contact experiments, where the apex of the finger is indented against a smooth glass plate, at constant normal force (see a sketch on Fig.2A). Using standard image analysis with a custom-made Matlab routine (see Supplementary Materials) we extract the centroid ($x_c,z_c$) and the contour of the cavity in polar coordinates $r(\theta,F_n)$, from which we deduce the static radial displacement $u^N_r(\theta,F_n) = r(\theta,F_n)-r(\theta,F_n=0)$ under normal load  (Figs.~2B and C). From elastic models for a cavity included in an incompressible elastic medium under an \emph{uniform} compressive load~\cite{jaeger2009fundamentals,sokolnikoff1956mathematical}, one would expect :
  \begin{equation}
    u_r(\theta) \approx  \frac{\sigma_0 a}{3E}.\left[\frac{3}{2} + 5 \cos(2 \theta) \right].
    \label{u3D}
  \end{equation}
In the present study, the cavity is subjected to the heterogeneous stress field generated by the Hertzian cylinder-on-plane-contact. However, as a first approach, we make the hypothesis that the compressive stress is homogeneous in the vicinity of the cavity and close to the value $\sigma_0$ derived from the Hertz theory. This assumption is supported by the relatively low values of $a/e\sim a/z\sim$ 0.2. Accordingly, the stress experienced by the cavity at a height $z$ from the apex of the cylinder writes
        \begin{equation}
        \label{sigma_cylindre}
            \sigma_0 = \frac{2F_{n}}{\pi e L \sqrt{1 + \frac{z^{2}}{e^{2}}}} , \mathrm{\quad yielding \quad} u_r\sim \alpha \sqrt{\frac{F_n}{1+\frac{\beta}{F_n}}}, 
        \end{equation}
\noindent where $\alpha = 2a/3\sqrt{1/(3\pi R L E)}$ and $\beta = z^2\pi L E/(3R)$ are two constants. The experimental measurements of $u_r$ are well adjusted by a $\cos{(2\theta+\varphi)}$ function (see Fig.~2D), where $\varphi \sim \pi/8$ is a phase shift required to adjust the data,due to the fact that the cavity is not perfectly located in the contact symmetry plane (see Fig.~1~A4 and dedicated measurements in the Supplementary Materials). The amplitude of the $u_r$ modulations under normal load, $U_r^N$, are extracted from the standard deviation of the $u_r(\theta)$ curves, and are plotted against the normal force on Fig.\ref{Fig2}E.  The data are reasonably well fitted with Eq.\ref{sigma_cylindre}, with $\alpha = 11\pm2 \mu$m/$N^{1/2}$ and $\beta = 1.9\pm 1$N, the order of magnitude of which compares to the theoretical values $\alpha \approx 24\mu$m/$N^{1/2}$ and $\beta = 0.8$~N. The cavity’s response sensitivity $\chi$ to normal indentation can be calculated from Eq.2, it writes :
\begin{equation}
  \chi = \frac{1}{a}\left(\frac{\partial U_r^N}{\partial F_n}\right)_{F_n \rightarrow 0} = \frac{\alpha}{a \sqrt{\beta}}=\frac{2}{3\pi} \frac{1}{ELz} 
\label{chi}
\end{equation}

The sensitivity is expected to scale as $1/z$ and to diverge as the cavity’s location approaches the cylinder contact plane.\\

\begin{figure*}[t!]
\includegraphics[width=\textwidth]{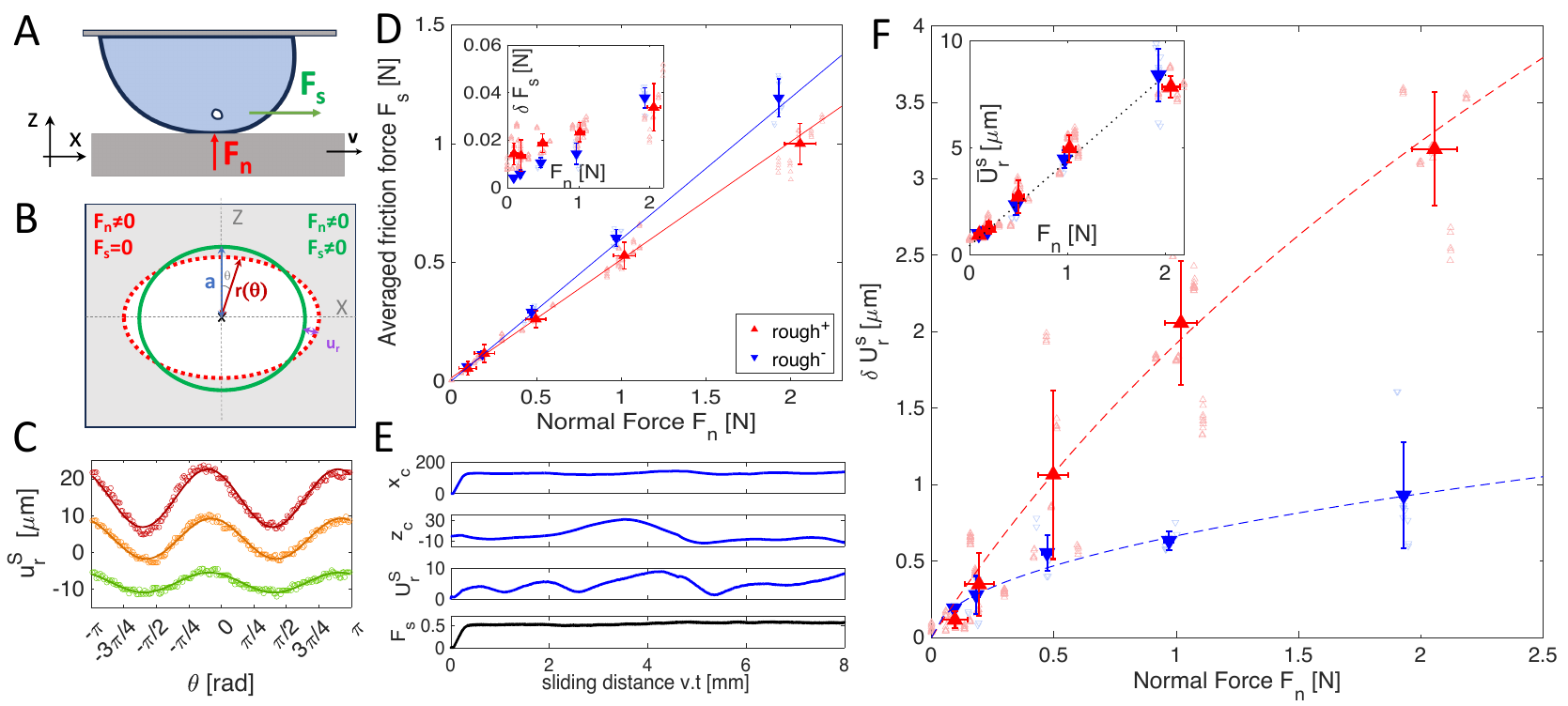}
\caption{\label{Fig3} A) Sketch of the semi-cylinder in steady sliding against a plane rough surface, at constant velocity $v$ and constant normal force $F_n$. B) Sketch of the unsheared (dotted line) and sheared (dashed) cavity, from which the radial displacement $u_r^s(\theta)$ is deduced. C) Selection of radial displacement $u_r^s(\theta)$ for the rough+ sample, at a given sliding distance ($d$=7.33~mm), and for normal forces $F_n = 0.5 N$ (lower curve), $F_n = 1 N$ (middle curve), $F_n = 2 N$ (upper curve). D) Time averaged friction force $<F_s>$ as a function of the normal force for the rough+ (upward triangles) and rough- (downward triangles) rough surfaces. Inset : friction force fluctuations $<\delta F_s>$ as a function of $F_n$.  The typical radial displacement amplitude $U_r^s(t)$ at time $t$ is obtained as $\sqrt{2}$ times the standard deviation of $u_r^s(\theta)$ over $\theta$. E) Example of the four measured observables : ($x_c, z_c, U_r^s$, in $\mu$m) and $F_s$ in N as a function of the sliding distance for the case of the rough+ sample at $F_n=1N$. F) Time fluctuations of $U_r^s(t)$, denoted $\delta U_r^s$ as a function of $F_n$. The dashed lines are guides for the eyes. Inset : time averaged $\bar{U_r^s}$ as a function of $F_n$. }
\end{figure*}
\subsection{Frictional response}
Secondly, we performed experiments where rough surfaces are slid at a constant velocity $v$ and constant normal force against the semi-cylinder (see a sketch on Fig.~3A), while measuring the friction force and the cavity deformations along time. We measured, at each time step in the steady sliding regime, the friction force, the centroid coordinates ($x_c,z_c$) of the cavity (see Fig. 3E) and its contour. In an attempt to isolate the shear component of the cavity deformation, we computed the radial displacement under friction force with respect to the static normal load radial profile, $u^S_r(\theta,F_n,t) = r(\theta,F_n,F_s\ne0)-r(\theta,F_n, F_s=0)$ (see Fig.~3B). We plot on Fig.~3C examples of these radial displacements under shear. Typically, the shear component induced by friction induces a sinusoidal modulation which tends to oppose to the normal load induced deformation, with a radial expansion at the poles and reduction at the equator. The typical amplitude of this signal $U^S_r$ is extracted.\\
For these four observables ($x_c(t)$, $z_c(t)$, $F_s(t)$, $U^S_r(t)$) we computed their time average in the steady sliding regime and their fluctuations ($\delta x_c$, $\delta z_c$, $\delta F_s$, $\delta U^S_r$, respectively). The associated measurement noises are estimated from out-of-contact experiment (see Supplementary Materials). As anticipated from the weak logarithmic dependence of the frictional stress of PDMS on velocity~\cite{nguyen2013,fazio2021} in the investigated velocity range, we found that none of these observables depend on the sliding velocity in the probed range ($v$ from 0.05 to 0.2 mm/s, see Supplementary Materials), indicating that the process is quasi-static. We thus combined results at different velocities. \\
On Fig.~3D we plot the time averaged friction force as a function of the normal load, and observe a linear relationships for both roughnesses. The dynamical friction coefficient $\mu =\bar{F_s}/F_n$ is slightly higher for the rough- texture. However, we found that the time fluctuations of the friction force $\delta F_s$ (inset of Fig.~3 D) are similar for both surfaces. We plot on the inset of Fig.~3F the time average radial displacement $\bar{U^S_r}$ as a function of $F_n$. A linear relationship is obtained for both rough samples, but here again, does not allow for texture discrimination. On the contrary, we find that the time fluctuations of $U^S_r$ are capable of discriminating textures, being three fold larger at $F_n$=2 N for the rough+ sample than for the rough- one (Fig.3F, main pannel).\\
 \section{Discussion}
Altogether, this work propose a mechanical framework to predict mechanoreceptor deformations upon stereotypical tactile tasks.
Is this simplified model applicable to natural mechanoreceptors ? The typical membrane tensions $\gamma$ triggering the response of biological mechanoreceptors lie around the mN/m scale. Taking a typical value of $E=$ 100 kPa for the Young's modulus of the human finger~\cite{opricsan2016experimental} one obtains an elastocapillary length~\cite{style2015stiffening} $l_{ec} = \gamma/E\sim 10$ nm, extremely small with respect to the typical mechanoreceptor diameter (tens to hundreds of $\mu$m). As a consequence, elastic stresses are dominant to predict the mechanoreceptor deformations.\\
Using this elastic model, we found that the static contact sensitivity $\chi$ increases and diverges as the cavity approaches the cylinder apex (Eq. \ref{chi}). However, from a biological perspective, a balance is likely required between tactile sensitivity and the structural integrity of mechanoreceptors during tactile exploration. Merkel corpuscles, specialized touch mechanoreceptors responsible for detecting static contact, are found approximately 1 mm beneath the skin surface, within dermal papillae (see, for example, \cite{deflorio2022skin}). These corpuscles are situated just below the epidermis, which may possibly act as a protective barrier.
\begin{figure*}[t]
\includegraphics[width=0.9\textwidth]{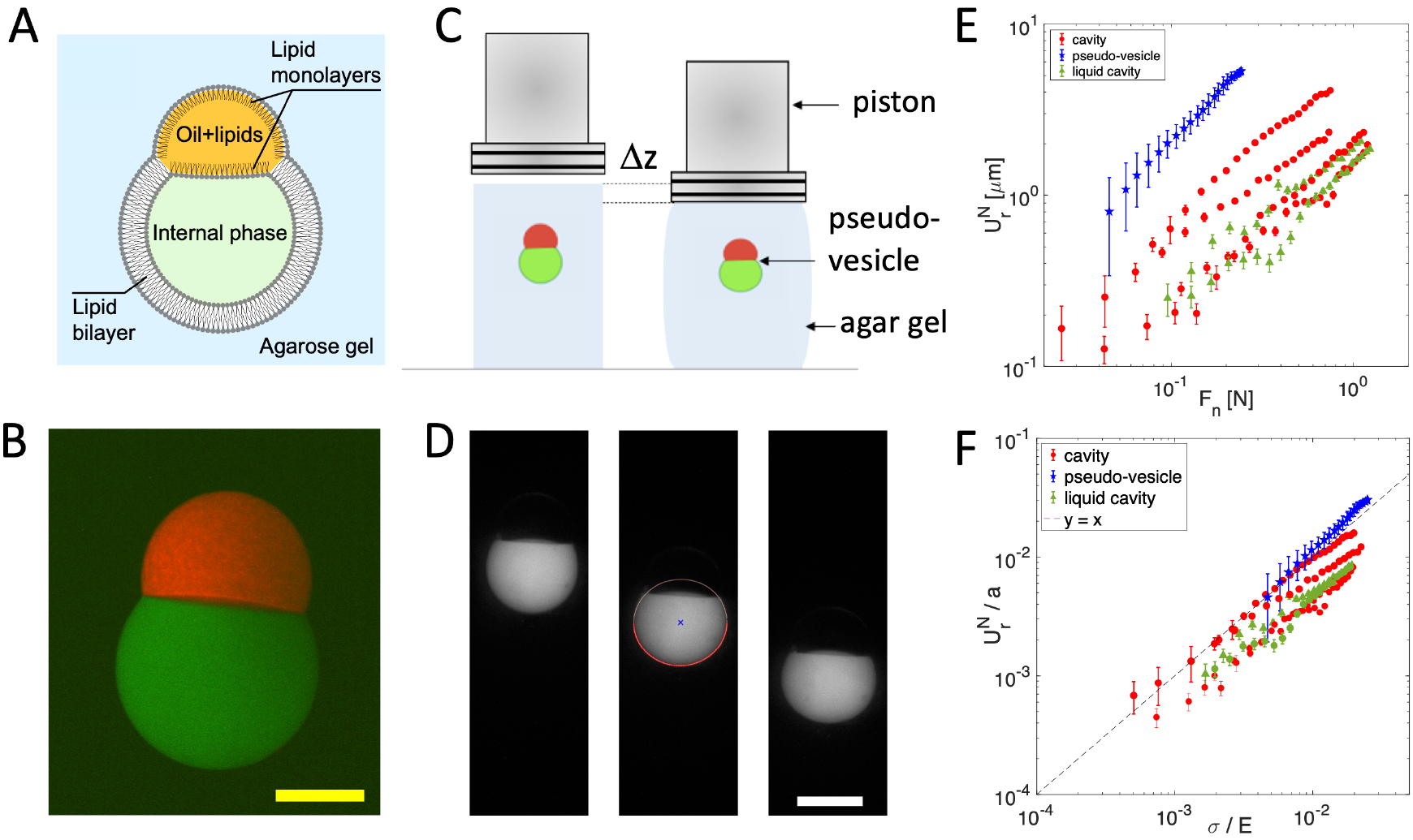}
\caption{See also \cite{tapie2023simple} A) Sketch of a pseudo-vesicle trapped in an agar gel. B) Fluorescence image of the pseudo vesicle, whose internal aqueous phase is marked with a green fluorophore and the oil phase in red (scale bar 200 $\mu$m). C) Sketch of the plane/plane deformation setup. D) Fluorescence images of the pseudo-vesicle upon deformation. The scale bar is 400 $\mu$m long. E) Radial deformation $U_R^N$ as a function of the normal force, for gas cavities (red discs) a liquid cavities (green triangles) and pseudo-vesicle (blue stars). F) Rescaled $U_r^N / a$ as a function of rescaled stress $\sigma/E$.}
\label{Fig4}
\end{figure*} 
\par In addition, natural mechanoreceptors embed a complex biological aqueous content within a plasma membrane, made of a lipid bilayer in which various membrane proteins are inserted. To add an extra layer of complexity, the external extracellular matrix is also viscoelastic. To what extent do these properties modify the mechanical response of natural mechanoreceptors ? To take a step closer to replicating natural mechanoreceptors, we previously developed lipid vesicle-like systems embedded in a hydrogel. In a recent paper \cite{tapie2023simple}, we demonstrated how to create and trap a lipid pseudo-vesicle within an aqueous agar gel. The term \textit{pseudo-vesicle} refers to a lipid vesicle with an oily cap on top (Fig.~4A and B). \\ 
The agar gel can be indented using a piston with a cross-sectional area $S$ (Fig.~4C and D), and its deformation $u_r^N$ can be measured as a function of the normal load in a plane-plane contact. In the Supplementary Materials of \cite{tapie2023simple}, we showed that the deformation of the pseudo-vesicle closely approximates that of a gas cavity in a PDMS elastomer, as studied in the present work.  \\
Notably, when the deformation $U_r^N$ is rescaled by the radius $a$ of the pseudo-vesicle (or cavity), and the applied compressive stress $\sigma = F_n/S$ is normalized by the Young’s modulus $E$, the curves for different systems collapse onto a single trend (Fig.~4E and F).\\
These results constitute additional clues to support our modeling of mechanoreceptor deformations using the approximation of a gas cavity, as a reasonable first approach for predicting their behavior in typical tactile tasks.\\
\par Using this elastic model under static normal load, we can estimate the order of magnitude of the strain at the cavity wall as $\epsilon \approx \ln(1+U_r/a) \approx 0.02$. To this strain corresponds a stress $\sigma_{\theta \theta}\sim E u_r^N/a \sim\sigma_0$ of the order of 10~kPa. For biological mechanoreceptors, to what extent does the membrane tension $\Delta\gamma$ increases upon contact stress ? It can be estimated from the typical area expansion modulus $K\sim$ 100 mN/m of a cell membrane~\cite{najem2015activation,reddy2012effect}: $\Delta\gamma = K \frac{\Delta\mathcal{A}}{\mathcal{A}}$, with $\mathcal{A}$ the membrane surface. One can estimate $\Delta\gamma \sim U_r^2/a^2\sim K (\frac{\sigma_0}{E})^2$. With the aforementioned value of $E$ one finds $\Delta\gamma\sim$ 1 mN/m, a value in the range of opening threshold tensions of mechanosensitive proteins\cite{ranade2014piezo2,delmas2022piezo}.\\
Secondly, we notice from Eq.~\ref{sigma_cylindre} that the anistropic experimental radial displacement $U_r^N$ yields to a theoretically anisotropic hoop stress $\sigma_{\theta \theta}$, being compressive at the cavity equator and extensional at the poles. This observation is known in the context of geophysical studies~\cite{lajtai1971theoretical,davis2017stress} to explain the localization of the fracture plane in gas cavities embedded in rocks. Such a stress anisotropy has never been mentioned in the context of touch mechanoreceptors. Back to the biological problem, this suggests that depending on their angular position on the membrane, the mechanosensitive protein pores will not have the same probability of opening. One may further wonder whether the density of mechanosensitive proteins in the mechanoreceptor membrane is isotropic, or whether Evolution has favored an anisotropic distribution (at the poles and not at the equator) to maximize tactile sensitivity. A second - more likely - hypothesis would be that the angular distribution of protein density changes in response to the local membrane curvature by a negative curvotaxis-like process, as reported for many different membrane proteins~\cite{has2021recent,johnson2024protein,sorre2012nature,prevost2015irsp53}. \\
Considering texture discriminations, our findings suggest that shape fluctuations, and as a consequence membrane stress fluctuations, allows for roughness discrimination of probed surfaces. This study calls for mechanical models of cavities under friction forces to assess this experimental finding. A first step in this direction is to measure and model the cavity's response to the passage of an elementary texture (a single defect), as we outline in the Supplementary Materials. On the biological side, our results brings a mechanical reason of being for Fast Adapting mechanoreceptors, whose neural response is sensitive to stress fluctuations.

\end{document}